\documentclass[12pt]{article}
\title{Much ado about 248}

\author{M.C. Nucci and P.G.L. Leach\footnote{permanent address: School of
Mathematical Sciences, Westville Campus, University of
KwaZulu-Natal, Durban 4000, Republic of South Africa}}
\date{Dipartimento di Matematica e Informatica,
Universit\`a di Perugia, 06123 Perugia, Italy}

\begin{document}

\maketitle

\begin{abstract}
In this note we present three representations of a 248-dimensional
Lie algebra, namely the algebra of Lie point symmetries admitted
by a system of five trivial ordinary differential equations each
of order forty-four, that admitted by a system of seven trivial
ordinary differential equations each of order twenty-eight and
that admitted by one trivial
 ordinary differential equation of order two hundred and forty-four.
\end{abstract}

\section{Introduction}

A system of $n$ ordinary differential  equations
each of order $M>1$,
\begin{equation}
u_k^{(M)}=f_k(u_j^{(s)},t),\;\;\; j,k=1,n,\;\;\; s=0,M-1, \label{1.1}
\end{equation}
has a variable number of Lie point symmetries depending upon the structure of the functions $f_k$.
The maximal dimension $D$ of the algebra of admitted Lie point symmetries can be obtained
  by the formul\ae\ \cite{Gonsalez-Gascon 84}
\begin{eqnarray}
 M=2 &\Longrightarrow& D=n^2+4n+3 \label{1.2} \\
 M>2 &\Longrightarrow & D=n^2+M n+3. \label{1.3}
\end{eqnarray}
Some explicit numbers are given in Table \ref{tab 1}.
\begin{table}
\caption{\label{tab 1}:  The maximal dimension of the algebra of
admitted Lie point symmetries for systems of equations of varying
order (horizontal) and number (vertical).}
\begin{center}
\begin{tabular}[h]
{|c||c|c|c|c|c|c|c|c|c|}
\hline $_n{\rm \backslash}^{{\mbox{\tiny M}}}$ &2 & 3& 4 & 5 & 6 &
7 & 8 & 9 & 10
\\ \hline \hline 1& 8& 7 & 8 &9 & 10 & 11 & 12 & 13 & 14\\ \hline 2& 15& 13 & 15 & 17
& 19 & 21 & 23 & 25 & 27 \\ \hline 3& 24& 21 & 24 & 27 & 30 & 33 &
36 & 39 & 42
\\ \hline 4 & 35 & 31 & 35 & 39 & 43 &47 & 51 & 55 & 59\\ \hline 5& 48 &43 & 48& 53 &
58 & 63 & 68 & 73 & 78
\\ \hline 6& 63 & 57 & 63 & 69 & 75 & 81 & 87 & 93 & 99
\\ \hline 7& 80 & 73 & 80 & 87 & 94 & 101 & 108 & 115 & 122
\\ \hline 8& 99 & 91 & 99 & 107 & 115 & 123 & 131 & 139 & 147
\\ \hline 9& 120 & 111 & 120 & 129 & 138 & 147 & 156 & 165 & 174
\\ \hline 10 & 143 & 133 & 143 & 153 & 163 & 173 & 183 & 193 & 203
\\ \hline
\end{tabular}
\end{center}
\end{table}

Recently the elaboration of the elements of the Lie algebra, $E8$,
of order 248 has been variously announced \cite {BBC, Derspiegel,
Lemonde, TimesLon, TimesNY} in the serious popular media.  The
authoritative source is the Atlas of Lie Groups and
Representations \cite{atlas} which is funded by the National
Science Foundation through the American Institute of Mathematics
\cite{institute}.  The results of the E8 computation were
announced in a talk at MIT by David Vogan on Monday, March 19,
2007, and the details may be found at \cite{technical}.  The Atlas
of Lie Groups and Representations is a project to make available
information about representations of semisimple Lie groups over
real and $p$-adic fields. Of particular importance is the problem
of the unitary dual, {\it ie} the classification of all of the
irreducible unitary representations of a given Lie group.  The
goal of the Atlas of Lie Groups and Representations is to classify
the unitary dual of a real Lie group, $G$, by computer.  A step in
this direction is to compute the admissible representations of $G$
including their Kazhdan-Lusztig-Vogan polynomials.  The
computation for $E8$ was an important test of the technology.
While the computation is an impressive achievement, it is only a
small step towards the unitary dual and should not be ranked as
important as the original work of Kazhdan, Lusztig, Vogan,
Beilinson, Bernstein {\it et al}. (See for example \cite{Beilinson
83 a, Beilinson 81 a, Bernstein 86 a, Kazhdan 79 a, Kazhdan 80 a,
Lusztig 83 a, Vogan 83 a, Gelfand 82 a}.) Nevertheless the result
was regarded as being suitable for a concerted campaign of
publicity to heighten awareness of Mathematics in the community at
large:\\
``Symmetrie ist m\"oglicherweise das erfolgreichste Prinzip der
Physik \"uberhaupt" \cite{Derspiegel}.\\
``Un groupe de chercheurs am\'ericains et europ\'eens, parmi
lesquels on trouve deux Fran\c{c}ais, est parvenu \`a d\'ecoder
une des structures les plus vastes de l'histoire des
math\'ematiques" \cite{Lemonde}. \\
``It may be that some day this calculation can help physicists to
understand the universe" \cite{TimesLon}.\\
``Eighteen mathematicians spent four years and 77 hours of
supercomputer computation to describe this structure"
\cite{TimesNY}.\\

In this note we demonstrate three representations of a Lie algebra
of dimension 248. The two of us spent four hours and  77 seconds
of pocket-calculator computation to describe these three
structures.

\section{Three simple systems}

For $D=248$ formula (\ref{1.2}) does not have integral solutions
and so there is no system of second-order ordinary differential
equations of maximal symmetry possessing a 248-dimensional algebra
of its Lie point symmetries\footnote{Is this another instance of
the intrinsically uniqueness of Classical Mechanics?}. About
formula (\ref{1.3}) the factors of 248-3=245 are 1, 5 and 7 (49 is
out of question because $49^2>245$). Consequently possible values
of $n$ are 1, 5 and 7. The corresponding values of $M$ are 244, 44
and 28, respectively. The systems of maximal symmetry are easily
obtained as one simply puts $f_k = 0$ $\forall$ $k$. Thus the
systems we construct are the simplest representations of the
equivalence class under point transformation of systems of
equations of maximal symmetry.

Firstly we consider the following system:
\begin{equation}
 u_k^{(44)}=0,\quad \quad k=1,5.
 \label{sys1}
\end{equation}
It is easy to show that this simple system admits a
248-dimensional algebra of its Lie point symmetries since $5^2+5\cdot
44+3=248$.  The algebra is generated by the operators
\begin{equation}
\begin{array}{lcl}
\Gamma_1&=&t^2\partial_t+43 t\sum_{i=1}^{5} u_i\partial_{u_i},\\
[0.2cm] \Gamma_2&=&t\partial_t, \\ [0.2cm]\Gamma_3&=&\partial_t,\\
[0.2cm] \Gamma_{i,k}&=&u_{k}\partial_{u_i},\quad  k=1,5,\;
i=1,5\\ [0.2cm] \Gamma_{i+5,s}&=&t^s\partial_{u_i},\quad
s=0,43,\; i=1,5.
\end{array}
\end{equation}
Secondly we consider the system
\begin{equation}
 u_r^{(28)}=0,\quad \quad r=1,7.
 \label{sys2}
\end{equation}
This equally simple system admits a
248-dimensional algebra ($7^2+7\cdot
28+3=248$) of its Lie point symmetries generated by
\begin{equation}
\begin{array}{lcl}
\Gamma_1&=&t^2\partial_t+27 t\sum_{j=1}^{7} u_j\partial_{u_j},\\
[0.2cm] \Gamma_2&=&t\partial_t, \\
[0.2cm]\Gamma_3&=&\partial_t,\\
[0.2cm] \Gamma_{j,r}&=&u_{r}\partial_{u_j},\quad  r=1,7,\;
j=1,7\\
[0.2cm] \Gamma_{j+7,n}&=&t^n\partial_{u_j},\quad
n=0,27,\; j=1,7.
\end{array}
\end{equation}
Thirdly and finally the scalar equation,
\begin{equation}
 u^{(244)}=0,
 \label{sys3}
\end{equation}
admits a 248-dimensional Lie algebra ($1^2+1\cdot
244+3=248$) of its point symmetries generated by the operators
\begin{equation}
\begin{array}{lcl}
\Gamma_1&=&t^2\partial_t+243 t u\partial_{u},\\
[0.2cm] \Gamma_2&=&t\partial_t, \\ [0.2cm]\Gamma_3&=&\partial_t,\\
[0.2cm] \Gamma_4&=&u\partial_{u},\\
[0.2cm] \Gamma_{n+5}&=&t^n\partial_{u},\quad n=0,243.
\end{array}
\end{equation}

\section{Conclusion}

We have demonstrated three representations of Lie algebras of
dimension 248 which is the dimension of $E8$.  Although the
algebras we present are not simple, their method of construction
is.  The reason for this simplicity is that we used
representations for systems of equations of maximal symmetry.  We
do not deny that larger systems, be that in order or number, of
less than maximal symmetry could possibly have an algebra of
dimension 248, but even on the assumption that such systems be
linear the complexity of the calculation becomes immense
\cite{Gorringe 88 a} and defeats the purpose of the present note.

Note that we have used the simplest forms for the generators of
the algebras of the three systems, (\ref{sys1}), (\ref{sys2}) and
(\ref{sys3}), for our primary interest is the demonstration of the
existence of the algebras.  Normally one would use combinations
which reflect subalgebraic structures.  For example in the case of
(\ref{sys3}) for which the algebra is obviously $sl(250,I\!\!R)$
one would replace $\Gamma_2$ with ${\tilde\Gamma}_2 = 2t\partial_t
+ 243u\partial_u$ to underline the subalgebraic structure
$\{sl(2,I\!\!R)\oplus A_1\}\oplus_s 244 A_1$, where $\Gamma_1$,
${\tilde\Gamma}_2$ and $\Gamma_3$ constitute a representation of
$sl(2,I\!\!R)$, $\Gamma_4$ reflects the homogeneity of the
equation in the dependent variable and the 244-element abelian
subalgebra is composed of the solution symmetries, so called
because the coefficient functions are solutions of (\ref{sys3}).

\section*{Acknowledgements}

PGLL thanks the University of Kwazulu-Natal for its continued support.

\end{document}